\documentclass[preprint,showkeys,nofootinbib]{revtex4}

\usepackage[dvips]{graphicx,color}
\usepackage{epsfig}
\usepackage{amsmath}
\usepackage{amssymb}
\usepackage{booktabs}
\usepackage[dvipdfm,bookmarks=true,bookmarksnumbered=true,bookmarkstype=toc]{hyperref} 
\usepackage{here}

\allowdisplaybreaks[4]
\newcommand{\lsim}{\raise0.3ex\hbox{$\;<$\kern-0.75em\raise-1.1ex\hbox{$\sim\;$}}}
\newcommand{\gsim}{\raise0.3ex\hbox{$\;>$\kern-0.75em\raise-1.1ex\hbox{$\sim\;$}}}

\newcommand{\uR}{\tilde{u}_R}

\newcommand{\dR}{\tilde{d}_R}

\newcommand{\nuL}{\tilde{\nu}_L}

\begin{document}

%%%%% declaration for front matter%%%%%%%%%%%%%%%%%%%%%%%%%%%%%%%%%%%
\title{Constraints from Unrealistic Vacua in the Next-to-Minimal Supersymmetric Standard Model}

\author{Yoshimi Kanehata}
\email{yoshimi@cc.kyoto-su.ac.jp}
%\email{yoshimi@hep.phys.ocha.ac.jp}
\affiliation{Maskawa Institute for Science and Culture, Kyoto Sangyo University, Kyoto 603-8555, Japan}
\affiliation{Department of Physics, Ochanomizu University, Tokyo 112-8610, Japan}

\author{Tatsuo Kobayashi}
\email{kobayash@gauge.scphys.kyoto-u.ac.jp}
\affiliation{Department of Physics, Kyoto University, Kyoto 606-8502, Japan}

\author{Yasufumi Konishi}
\email{konishi@cc.kyoto-su.ac.jp}
\affiliation{Maskawa Institute for Science and Culture, Kyoto Sangyo University, Kyoto 603-8555, Japan}

\author{Osamu Seto}
%\email{osamu@hgu.jp}
\email{seto@physics.umn.edu}
\affiliation{Department of Architecture and Building Engineering, Hokkai-Gakuen University, Sapporo 062-8605, Japan}

\author{Takashi Shimomura}
\email{stakashi@yukawa.kyoto-u.ac.jp}
\affiliation{Yukawa Institute for Theoretical Physics, Kyoto University, Kyoto 606-8502, Japan}

\date{\today}

\preprint{YITP-11-25}
\preprint{KUNS-2323}
\preprint{MISC-2011-03}
\preprint{OCHA-PP-306}
\preprint{HGU-CAP 010}

%%%%% typeset front matter (including abstract) %%%%%%%%%%%%%%%%%%%%%%
\begin{abstract}
We study constraints to avoid deep unrealistic minima in the next-to-minimal supersymmetric standard model. We analyze a scalar 
potential along directions where all of and one of the three Higgs fields develop their vacuum expectation values, 
and find unrealistic minima deeper than the electroweak symmetry breaking (EWSB) vacuum. These unrealistic 
minima threaten the realization of the successful EWSB and therefore should be avoided. Necessary 
conditions to avoid these minima result in constraints of parameters. We show that a wide and significant region of 
the parameter space, especially large $\lambda$, is ruled out by our constraints. 
\end{abstract}

\maketitle

%%%%%%%%%%%%%%%%%%%%%%
\section{Introduction}\label{sec:introduction}
%%%%%%%%%%%%%%%%%%%%%%

The origin of the electroweak (EW) scale and its stability against large radiative corrections are unanswered 
questions in the standard model (SM) of elementary particles. New physics beyond the SM should provide a solution 
to stabilize its scale.  Supersymmetry (SUSY) is one of the promising frameworks in this regard. 
The simplest supersymmetric extension of the SM is the minimal supersymmetric standard model (MSSM).

It is well known that the MSSM suffers from a naturalness problem,
 the so-called $\mu$ problem \cite{Kim:1983dt}.
The MSSM has the supersymmetric Higgs/Higgsino mass term, $\mu$ term,  
with a dimensionful parameter $\mu$. Although $\mu$ is usually assumed to be of the 
order of the EW scale, there is no a priori reason for $\mu$ of the EW scale.
If the size of $\mu$ is much larger than the EW scale, new fine-tuning would be re-introduced
 to obtain the observed masses of gauge bosons.
On the other hand, if the size of $\mu$ is much smaller than the EW scale,
 it would conflict with non-observation of a new charged fermion, namely
 the charged Higgsino.

The next-to-minimal supersymmetric standard model (NMSSM) 
\cite{Fayet:1974pd,Fayet:1976et,Fayet:1977yc,Fayet:1979sa,Nilles:1982dy,Frere:1983ag,Derendinger:1983bz,Ellis:1988er,Drees:1988fc}
( for a review, see \cite{Ellwanger:2009dp}) 
is the simplest extension of the MSSM by adding 
one gauge singlet superfield $\hat S$ and $\mathbb{Z}_3$ symmetry.
The $\mu$ term is forbidden by the $\mathbb{Z}_3$ symmetry, instead is generated effectively after 
the singlet scalar $S$ develops its vacuum expectation value (vev).
The singlet scalar $S$ develops the vev associated
 with the standard radiative symmetry breaking at the EW scale.
Hence, its vev takes automatically a value of the order of EW scale.
Another advantage of the NMSSM is concerned about the Higgs mass.
{}From the non-observation of the CP-even light Higgs boson $h$ at LEP II, 
the lower bound on its mass $m_h$ has been obtained as $m_h > 114$ GeV \cite{Barate:2003sz}.
Within the MSSM, the lightest CP-even Higgs mass 
is bounded by the $Z$ boson mass $m_Z$ at the tree
level.
Radiative corrections due to stop (a superpartner of top quark) masses would increase the 
Higgs mass \cite{Okada:1990vk,Haber:1990aw,Ellis:1991zd}, but one may need a large stop mass such as ${\cal O}(1)$
TeV to make the lightest Higgs heavy enough above the LEP bound.
Such heavy stops would lead to a fine-tuning problem,
 in other words the little hierarchy problem 
\cite{Barbieri:1987fn,Chankowski:1997zh,Chankowski:1998xv,Kane:1998im,BasteroGil:1999gu,Kane:2002ap}\footnote{
See also e.g. 
\cite{Agashe:1997kn,Brignole:2003cm,Casas:2004gh,Kitano:2005wc,Choi:2005hd,Choi:2006xb,Dermisek:2006ey,Abe:2007kf,Horton:2009ed,Kobayashi:2010ye}.
}, 
which could be moderated 
in the NMSSM.
The Higgs potential of the NMSSM has a new quartic term, which 
increases the lightest CP-even Higgs mass at the tree level.

While the NMSSM has appealing features mentioned above,
 the structure of the scalar potential of the NMSSM is more complicated than that of the MSSM
 due to the presence of the additional singlet scalar $S$.
As far as the NMSSM scalar potential is concerned,
 a cosmological domain wall problem caused by the symmetries of the NMSSM
 has been well studied \cite{Ellis:1986mq,Gelmini:1988sf}.
The domain wall problem can be avoided by introducing suitable non-renormalizable operators
 that do not generate a dangerously large tadpole \cite{Abel:1996cr,Kolda:1998rm,Panagiotakopoulos:1998yw}.
However, this is not the end of story. 
Since the potential is complicated, phenomenologically unacceptable vacua could exist.  
Hence, it is important to analyze the vacuum structure of 
the Higgs scalar potential, and to derive conditions to avoid unrealistic minima 
and realize the successful EW symmetry breaking (EWSB).
These studies would provide us with significant constraints on 
the parameter space of couplings and soft SUSY breaking terms 
in the NMSSM.

In addition to the Higgs scalar fields, there are other scalar fields, 
which are superpartners of quarks and leptons, i.e. the so-called squarks and sleptons.
They might develop vevs and lead to unrealistic vacua.
On such vacua the colour and/or charge breaking (CCB) may 
occur if squarks and sleptons develop vevs \cite{Frere:1983ag,AlvarezGaume:1983gj,Derendinger:1983bz,
Kounnas:1983td,Claudson:1983et,Drees:1985ie,Gunion:1987qv,Komatsu:1988mt,Gamberini:1989jw}.
Moreover, the potential may include directions unbounded from
below (UFB). Systematic studies on such unrealistic vacua and UFB directions 
have been carried out in the MSSM  \cite{Casas:1995pd}.
Recently, such analyses were also extended including 
terms generating non-vanishing neutrino masses and 
the corresponding soft SUSY breaking terms \cite{Kobayashi:2010zx,Kanehata:2010ci}, and flavour physics, 
e.g. \cite{Casas:1996de,Park:2010wf,Hisano:2010re}. 
These analyses show that one has theoretical constraints 
among couplings and soft SUSY breaking terms in order to avoid 
unrealistic vacua and realize the successful EWSB.
Those constraints are useful to eliminate parameter space of the models, and hence are important.
Our purpose of this paper is to extend such systematic analyses on unrealistic vacua to the NMSSM.

In this paper, we analyze the vacuum structure of the Higgs scalar potential in the NMSSM.
We study unrealistic vacua and derive conditions to 
avoid them.
We investigate implications of those constraints by 
using examples of numerical analyses and 
simplifying the constraints in a certain limit.
We also study unrealistic vacua of scalar potential including 
squarks and sleptons.

This paper is organized as follows.
In section \ref{sec:real-vacu-nmssm}, we review the Higgs potential, the EWSB vacuum and Higgs masses of 
the NMSSM.
In section \ref{sec:constr-from-vacua}, we study unrealistic minima 
and show the conditions to avoid them.
In section \ref{sec:numerical-analysis}, we study our constraints numerically 
for several examples.
Section \ref{sec:summary-discussion} is devoted to conclusion and discussion.
We give our notations and the scalar potential of 
the NMSSM in Appendix \ref{sec:scalar-potential}.
In Appendix \ref{sec:ccb-constr-nussm}, we give detailed studies on 
the unrealistic minima, where squarks and sleptons 
as well as the singlet $S$ develop their vevs 
in the NMSSM.

%%%%%%%%%%%%%%%%%%%%%%%%%%%%%%%%%%%%%%%
\section{Realistic Vacuum of the NMSSM}\label{sec:real-vacu-nmssm}
%%%%%%%%%%%%%%%%%%%%%%%%%%%%%%%%%%%%%%%
The NMSSM is an extension of the MSSM by adding an extra gauge singlet scalar, $S$, and its fermionic partner, $\tilde{S}$. 
This new scalar participates in the EWSB together with two doublet 
Higgs scalars by developing their vevs. 
In this section, we briefly review a realistic vacuum of the NMSSM on which the EWSB successfully occurs. 
Notations of particles are summarized in Appendix \ref{sec:scalar-potential}.

The superpotential of the NMSSM is given by
\begin{align}
 \mathcal{W}_{\mathrm{NMSSM}} &= Y_d \hat{H}_1\cdot \hat{Q} \hat{D}^c_R 
                              + Y_u \hat{H}_2\cdot \hat{Q} \hat{U}^c_R 
                              + Y_e \hat{H}_1\cdot \hat{L} \hat{E}^c_R 
                              - \lambda\hat{S} \hat{H}_1\cdot\hat{H}_2 
                              + \frac{1}{3}\kappa\hat{S}^3, \label{eq:1}
\end{align}
where $Y_u,~Y_d$ and $Y_e$ are the Yukawa couplings of up-type quarks, down-type quarks and charged leptons, respectively, and 
$\lambda$ and $\kappa$ are Yukawa coupling constants of the Higgs scalars. Here, we impose a global 
$\mathbb{Z}_3$ symmetry to forbid tadpole and quadratic terms. The soft SUSY breaking terms are given by
\begin{align}
  V_{\mathrm{soft}} & = m^2_{H_1} H_1^\dagger H_1 + m^2_{H_2} H_2^\dagger H_2 + m^2_{S} S^\dagger S 
                      - \left( \lambda A_\lambda S H_1 \cdot H_2 
                       - \frac{1}{3} \kappa A_\kappa S^3 + h.c.\right) \nonumber \\
                    &~~+ m^2_{\tilde{Q}} \tilde{Q}^\dagger \tilde{Q} + m^2_{\tilde{u}_R} \tilde{u}_R^\ast \tilde{u}_R
                      + m^2_{\tilde{d}_R} \tilde{d}_R^\ast \tilde{d}_R 
                      + m^2_{\tilde{L}} \tilde{L}^\dagger \tilde{L} + m^2_{\tilde{e}_R} \tilde{e}_R^\ast \tilde{e}_R  \nonumber \\
                    &~~+ \big(A_d Y_d H_1 \cdot \tilde{Q} \tilde{d}_R^\ast + A_u Y_u H_2 \cdot \tilde{Q} \tilde{u}_R^\ast 
                      + A_e Y_e H_1 \cdot \tilde{L} \tilde{e}_R^\ast + h.c.\big), \label{eq:2}
\end{align}
where we assume that all of soft masses, trilinear couplings and Yukawa couplings are real for simplicity. Indices for 
generation of squarks and sleptons are omitted.
The scalar potential of the Higgs scalars can be obtained from $F$-,
$D$-terms given in Appendix \ref{sec:scalar-potential} and the soft SUSY breaking terms.
For the EW symmetry to be successfully broken, the neutral Higgs fields 
develop vevs while vevs of the charged Higgs fields are vanishing. 
Using the gauge transformations, without loss of generality, one can take $\langle H_2^+ \rangle = 0$ and 
$\langle H_2^0 \rangle = v_2 \in \mathbb{R}^+$. The condition for vanishing $\langle H_1^- \rangle$ is to require that 
the charged Higgs scalars have positive masses squared. Then, the Higgs potential of the neutral components is given by,
\begin{align}
 V 
=& \lambda^2|S|^2 \left(|H_1^0|^2 + |H_2^0|^2 \right) + \lambda^2 | H_1^0|^2 | H_2^0|^2 
 + \kappa^2 |S|^4 -  (\lambda \kappa H_1^0 H_2^0 (S^\ast)^2 + h.c.) \nonumber \\
& + \frac{1}{8} g^2 \left(|H_1^0|^2 - |H_2^0|^2 \right)^2 
 + m_{H_1}^2 |H_1^0|^2 + m_{H_2}^2 |H_2^0|^2 + m_{S}^2 |S|^2 \nonumber \\
& -  (\lambda A_\lambda H_1^0 H_2^0 S - \frac{1}{3} \kappa A_\kappa S^3 + h.c.), \label{eq:3}
\end{align}
where $g^2 = g_1^2 + g_2^2$, and $g_1$ and $g_2$ denote the gauge coupling constant of $U(1)$ and $SU(2)$, respectively.
The Higgs sector is characterized by the following parameters,
\begin{align}
\lambda, ~~\kappa, ~~m_{H_1}^2, ~~m_{H_2}^2, ~~m_{S}^2, 
~~A_\lambda ~~{\rm and} ~~A_\kappa.
\label{eq:parameter-1}
\end{align}
The remaining vevs of $H_1^0$ and $S$
in general can be complex. However, in \cite{Romao:1986jy}, it was shown that such CP violating extrema are maxima 
rather than minima. Thus, it is reasonable to assume that 
the neutral Higgs fields develop real and non-vanishing vevs and the charged ones do not. Then, we note vevs as 
\begin{align}
\langle H_1^0 \rangle = v_1, \qquad \langle H_2^0 \rangle = v_2, 
\qquad \langle S \rangle = s.
\end{align}
Furthermore, as was discussed in \cite{Cerdeno:2004xw}, the Higgs potential, (\ref{eq:3}), is invariant under the replacements, 
$\lambda,~\kappa,~s \rightarrow -\lambda,~-\kappa,~-s$ and $\lambda,~v_1 \rightarrow -\lambda,~-v_1$, 
therefore we can take $\lambda$ and $v_1$ to be always positive while $\kappa,~\mu (\equiv \lambda s)$ and 
$A_\lambda,~A_\kappa$ can have both signs. 
These vevs are determined by minimizing the potential, (\ref{eq:3}), 
with respect to the neutral Higgs scalars, that is, they satisfy 
the following stationary conditions,
\begin{subequations}\label{eq:19}
 \begin{align}
   \frac{\partial V}{\partial H^0_1} &= \lambda^2 v \cos\beta ( s^2 + v^2 \sin^2 \beta )
     - \lambda \kappa v s^2 \sin\beta + \frac{1}{4} g^2 v^3 \cos\beta \cos2 \beta \nonumber \\
    &\quad + m_{H_1}^2 v \cos\beta - \lambda A_\lambda v s \sin\beta = 0, \label{eq:20} \\
   \frac{\partial V}{\partial H^0_2} &= \lambda^2 v \sin\beta ( s^2 + v^2 \cos^2 \beta )
     - \lambda \kappa v s^2 \cos\beta - \frac{1}{4} g^2 v^3 \sin\beta \cos2 \beta \nonumber \\
    &\quad + m_{H_2}^2 v \sin\beta - \lambda A_\lambda v s \cos\beta = 0, \label{eq:21} \\
   \frac{\partial V}{\partial S} &= \lambda^2 s v^2 + 2 \kappa^2 s^3 - \lambda \kappa v^2 s \sin 2\beta
    + m_S^2 s - \frac{1}{2} \lambda A_\lambda v^2 \sin 2\beta + \kappa A_\kappa s^2 = 0, \label{eq:22}
 \end{align}
%\label{eq:19}
\end{subequations}
where $v = \sqrt{v_1^2 + v_2^2}$ and $\tan\beta = v_2/v_1$.

Here, let us classify solutions of the stationary conditions, 
(\ref{eq:19}). 
It is important to emphasize here that, without very special relations among parameters, 
when two of $v_1, v_2$ and $s$ are non-vanishing, the other 
must be non-vanishing, too.
This fact is originated from the trilinear terms, $\lambda A_\lambda H_1^0 H_2^0 S$, 
in the soft SUSY breaking terms and the quartic term, $\lambda \kappa H_1^0 H_2^0 (S^\ast)^2$, 
in the $F$-term potential.
Suppose that $v_1$ and $v_2$ are non-vanishing.
Then, (\ref{eq:22}) can not be satisfied for non-vanishing $A_\lambda$ unless $S \neq 0$.
Similar discussion is applicable to other cases. When we start with any combination 
of two non-vanishing vevs, we obtain the same result, that is, 
all of three vevs should be non-vanishing.
On the other hand, we find another type of the solutions that 
only one of  $v_1, v_2$ and $s$ is non-vanishing, while 
the others are vanishing.
Therefore, non-trivial solutions of (\ref{eq:19}) are either three Higgs fields are non-vanishing or one Higgs field is non-vanishing.
This observation justifies our strategy of analyses on unrealistic minima of the Higgs potential in the next section.

It is useful to express the soft SUSY breaking masses in terms of other parameters rewriting the stationary conditions, (\ref{eq:19}), 
\begin{subequations}\label{eq:4}
\begin{align}
m_{H_1}^2 &= - \mu^2 - \frac{2 \lambda^2}{g^2} m_Z^2 \sin^2\beta - \frac{1}{2} m_Z^2 \cos 2\beta 
           + \mu \left( \frac{\kappa}{\lambda} \mu + A_\lambda \right) \tan \beta, \label{eq:11}\\
m_{H_2}^2 &= - \mu^2 - \frac{2 \lambda^2}{g^2} m_Z^2 \cos^2\beta + \frac{1}{2} m_Z^2 \cos 2\beta 
           + \mu \left( \frac{\kappa}{\lambda} \mu + A_\lambda \right) \cot \beta, \label{eq:12}\\
    m_S^2 &= - \mu^2 - \frac{2 \kappa^2}{\lambda^2} \mu^2 + \frac{2 \lambda \kappa}{g^2} m_Z^2 \sin 2\beta 
           + \frac{\lambda^2}{g^2} \frac{A_\lambda m_Z^2}{\mu} \sin 2\beta - \frac{\kappa}{\lambda} A_\kappa \mu, \label{eq:13}
\end{align}%\label{eq:4}
\end{subequations}
where $m_Z^2 = \frac{1}{2} g^2 v^2$ and $\mu = \lambda s$.
Thus, given $m_Z$, we can use the following parameters,
\begin{align}
\lambda, ~~\kappa, ~~A_\lambda, ~~A_\kappa,~~\tan \beta 
~~{\rm and} ~~\mu,
\label{eq:parameter-2}
\end{align}
instead of (\ref{eq:parameter-1}).
Using these parameters, the realistic minimum of the potential, which reproduces the observed $Z$ boson mass, can be written as
\begin{align}
 V_\mathrm{min} &=
- \lambda^2 \frac{m_Z^4 \sin^2 2 \beta}{g^4} - \frac{m_Z^4 \cos^2 2 \beta}{2 g^2} 
+ \overline{V}^{S}_{\mathrm{min}}, \label{eq:6}
\end{align}
where $\overline{V}^S_{\mathrm{min}}$ is the potential involving only $s$ or $\mu/\lambda$, 
\begin{align}
 \overline{V}^S_{\mathrm{min}} = 
\frac{\kappa^2}{\lambda^4}  \mu^4
+ \frac{2}{3} \frac{\kappa}{\lambda^3} A_\kappa \mu^3
+ \frac{1}{\lambda^2} m_S^2 \mu^2, \label{eq:5}
\end{align}
with $m_S^2$ given by \eqref{eq:13}.
In the following sections, we study unrealistic or CCB vacua and 
compare their depths with (\ref{eq:6}).

If the vev of $s$ $(\mu/\lambda)$ is large enough and the potential, $\overline{V}^S_{\mathrm{min}}$, is dominant in the full potential, 
the typical depth of the realistic minimum can be estimated as 
\begin{align}
 V_{\mathrm{min}} \simeq \overline{V}^S_{\mathrm{min}} 
  \simeq - \frac{\kappa^2}{\lambda^4}\mu^4 - \frac{\kappa}{3 \lambda^3} A_\kappa \mu^3.\label{eq:31}
\end{align}
Such an approximation is useful to estimate constraints, which will be shown in the next section.

Before we move to analysis on the scalar potential, we show mass-squared matrices of the Higgs bosons in order to examine  
tachyonic masses in the next section. Degrees of 
freedom of the Higgs bosons are ten, and three of them are absorbed by gauge bosons via the Higgs mechanism. 
Remaining seven physical degrees correspond to three CP-even Higgs bosons, two CP-odd Higgs bosons and one charged Higgs boson. 
The mass-squared matrix of the CP-even Higgs bosons is real-symmetric and given by 
\begin{subequations}\label{eq:23}
 \begin{align}
  M_{h,11}^2 &= m_Z^2 \cos^2 \beta + \mu \left(\frac{\kappa}{\lambda} \mu + A_\lambda \right) \tan \beta , \\
  M_{h,22}^2 &= m_Z^2 \sin^2 \beta + \mu \left(\frac{\kappa}{\lambda} \mu + A_\lambda \right) \cot \beta , \\
  M_{h,33}^2 &= \frac{4 \kappa^2}{\lambda^2} \mu^2 + \frac{\kappa}{\lambda} A_\kappa \mu 
              + \frac{\lambda^2}{g^2} \frac{A_\lambda m_Z^2}{\mu} \sin 2\beta, \\
  M_{h,12}^2 &= 2 \left( \frac{\lambda^2}{g^2} - \frac{1}{4} \right) m_Z^2 \sin 2\beta 
              - \mu \left(\frac{\kappa}{\lambda} \mu + A_\lambda \right), \\
  M_{h,13}^2 &= \frac{2\sqrt{2} \lambda}{g} \mu m_Z \cos\beta 
              - \frac{\sqrt{2} \lambda}{g} m_Z \left( A_\lambda + \frac{2 \kappa}{\lambda} \mu \right) \sin\beta, \\
  M_{h,23}^2 &= \frac{2\sqrt{2} \lambda}{g} \mu m_Z \sin\beta 
              - \frac{\sqrt{2} \lambda}{g} m_Z \left( A_\lambda + \frac{2 \kappa}{\lambda} \mu \right) \cos\beta.
 \end{align}
\end{subequations}
The mass-squared matrix of the CP-odd Higgs bosons is also real-symmetric and given by 
\begin{subequations}\label{eq:24}
 \begin{align}
  M^2_{A,11} &= \frac{2 \mu}{\sin 2\beta} \left( A_\lambda + \frac{\kappa}{\lambda} \mu \right), \\
  M^2_{A,22} &= \frac{\lambda^2}{g^2} m_Z^2 \left( \frac{A_\lambda}{\mu} + \frac{4 \kappa}{\lambda} \right) \sin 2\beta 
              - \frac{3 \kappa}{\lambda} A_\kappa \mu, \\
  M^2_{A,12} &= \frac{\sqrt{2} \lambda}{g} m_Z \left( A_\lambda - \frac{2 \kappa}{\lambda} \mu \right).
 \end{align}
\end{subequations}
It can be understood in (\ref{eq:23}) and (\ref{eq:24}) that physical masses become tachyonic if $\lambda$ is large and hence 
the off-diagonal elements becomes comparable or larger than the diagonal ones.
The mass squared of the charged Higgs boson is 
\begin{align}
 m^2_{H^\pm} = m_W^2 - \frac{2 \lambda^2}{g^2} m_Z^2 + \frac{2 \mu}{\sin 2\beta} \left(A_\lambda + \frac{\kappa}{\lambda} \mu \right), \label{eq:25}
\end{align}
where $m_W^2 = \frac{1}{2}g_2^2 v^2$ is the mass squared of the $W$ boson. 
The charged Higgs boson mass squared can also be tachyonic when $\lambda$ is large enough.
These mass-squared matrices are used in numerical calculations to find tachyonic mass regions.

%%%%%%%%%%%%%%%%%%%%%%%%%%%%%%%%%%%%%%%%%%%%%%%%%%%%%%%%%%%%%%%%%%%%%%%%%%%%%%%
\section{Constraints from unrealistic and Colour and/or Charge Breaking minima}\label{sec:constr-from-vacua}
%%%%%%%%%%%%%%%%%%%%%%%%%%%%%%%%%%%%%%%%%%%%%%%%%%%%%%%%%%%%%%%%%%%%%%%%%%%%%%%
In this section, we show that unrealistic minima and/or CCB 
minima appear in the scalar potential and derive necessary conditions to avoid these minima. 
First we consider directions where the neutral Higgs fields are non-vanishing while 
all of other scalar fields are vanishing. Next we consider directions where squarks and/or sleptons as well as 
the Higgs fields are non-vanishing. In the following, we discuss directions involving only the neutral Higgs fields 
and note $H_{1,2}^0$ as $H_{1,2}$ for abbreviation. As discussed in the previous section,
when two of the Higgs fields develop their vevs, 
the other must  develop its vev to satisfy the stationary conditions.
Then, analyses of the scalar potential are constrained to cases of either one or three non-vanishing Higgs fields.
The realistic minimum given in the previous section 
is included along the direction, where all of three Higgs fields develop their vevs.
Such a direction with all of three non-vanishing Higgs vevs 
may include other unrealistic minima. 
Indeed, one unrealistic minimum with $|H_1| = |H_2| \neq 0$ and  $S\neq 0$ will 
be studied below.
However, analyses with three non-vanishing Higgs fields are so complicated in general 
that it can not be solved analytically. Hence we restrict our
discussions to four possible cases in which three Higgs fields are aligned as $|H_1| = |H_2| \neq 0$ and  $S\neq 0$ so that 
$D$-term and $F_S$-term are vanishing, and one of the three Higgs fields is non-vanishing.  
In fact, deeper minima than the realistic minimum are easily found 
along these directions. Such directions should be avoided to stabilize the realistic minimum. 
One of the main purpose of this paper is to show these directions are
really dangerous for realization of the realistic minimum.
Furthermore, along the directions with 
non-vanishing vevs of squarks and/or sleptons, we systematically study CCB directions and derive necessary conditions 
to avoid these CCB minima according to general properties for CCB directions discussed in \cite{Casas:1995pd}. 

%%%%%%%%%%%%%%%%%%%%%%%%%%%%%%%%%%%%%%%%%%%%%%%%%%%%%%%%%%%%%%%%%%%%%%%%%%%%%%%%%%%%%
\subsection{Unrealistic minimum along $|H_1| = |H_2| \neq 0$ and $S\neq 0$ direction}\label{sec:unre-minim-along-H1H2S}
%%%%%%%%%%%%%%%%%%%%%%%%%%%%%%%%%%%%%%%%%%%%%%%%%%%%%%%%%%%%%%%%%%%%%%%%%%%%%%%%%%%%%
We consider the direction that 
\begin{align}
 |H_1| = |H_2| \neq 0,\quad S \neq 0,
\end{align}
where the up-type Higgs and the down-type Higgs fields have the same vev.
This direction corresponds to the so-called MSSM UFB-1 direction 
with non-vanishing gauge singlet $S$.
Along this direction, the $D$-term vanishes.
Then, it would lead to a UFB direction in the MSSM without $S$.
However, in the present case with $S \neq 0$, the potential is lifted up at a large value of the gauge singlet scalar, thus a minimum appears. 
The scalar potential along this direction is given by
\begin{align}
 V^{H_1H_2S} &= 2 \lambda^2 |S|^2 |H_2|^2 + |F_S|^2 - ( \lambda A_\lambda S H_1 H_2 - \frac{1}{3}\kappa A_\kappa S^3 + h.c.) \nonumber \\ 
           &\quad +(m_{H_1}^2 + m_{H_2}^2)|H_2|^2 + m_S^2 |S|^2,
\end{align}
where
\begin{align}
 F_S = -\lambda H_1 H_2 + \kappa S^2. \label{eq:28}
\end{align}
The deepest minimum can be found along which trilinear couplings are negative and the $F_S$-term is vanishing, 
that is 
\begin{align}
 S^2 = \frac{\lambda}{\kappa} H_1 H_2.\label{eq:41}
\end{align}
Note that the parameter, $\kappa$, must be positive to satisfy this relation.
Then, the potential is written as
\begin{align}
 V^{H_1H_2S} = \hat F |H_2|^4 - 2 \hat A |H_2|^3 + \hat m^2 |H_2|^2, \label{eq:36}
\end{align}
where
\begin{subequations}
\begin{align}
 & \hat{F} = \frac{2 \lambda^3}{\kappa}, \label{eq:38} \\
 & \hat{A} = \lambda \sqrt{\frac{\lambda}{\kappa}} \left| A_\lambda - \frac{1}{3} A_\kappa \right|, \label{eq:39} \\
 & \hat{m}^2 = m_{H_1}^2 + m_{H_2}^2 +  \frac{\lambda}{\kappa} m_S^2. \label{eq:40}
\end{align}
\end{subequations}
The trilinear term, $\hat{A}$, can always be taken to be positive using the sign of $S$.
By minimizing the potential of (\ref{eq:36}) with respect to $|H_2|$, the value of $|H_2|$ at extremal 
is obtained as
\begin{align}
 |H_2|_{\mathrm{ext}} = \frac{3 \hat{A}}{4 \hat{F}}
                       \left( 1 + \sqrt{1 - \frac{8 \hat{m}^2 \hat{F}}{9 \hat{A}^2} }~ \right), \label{eq:26}
\end{align}
where $\hat{m}^2 \le \frac{9 \hat{A}^2}{8 \hat{F}}$ is required for $|H_2|_{\mathrm{ext}}$ to be real.
Then, the minimum of the potential is obtained by inserting (\ref{eq:26}) as
\begin{align}
 V^{H_1H_2S}_{\mathrm{min}} = -\frac{1}{2} |H_2|_{\mathrm{ext}}^2 (\hat{A} |H_2|_{\mathrm{ext}} - \hat{m}^2). \label{eq:27}
\end{align}
To realize the realistic minimum, the following condition is required,
\begin{align}
 V^{H_1H_2S}_{\mathrm{min}} \ge V_{\mathrm{min}}.\label{eq:16}
\end{align}

In the next section, we examine numerically this condition. Let us see intuitive implications of this condition by using 
some approximation. The extremal value is roughly estimated as 
\begin{align}
 |H_2|_{\mathrm{ext}} \simeq \frac{|A_\kappa - 3 A_\lambda|}{8} 
  \sqrt{\frac{\kappa}{\lambda^3}},\label{eq:34}
\end{align}
and the depth of the minimum is 
\begin{align}
 V^{H_1 H_2 S}_{\mathrm{min}} \simeq -\frac{27}{1024} \frac{\kappa}{\lambda^3} A_\lambda^4, \label{eq:37}
\end{align}
for $A_\lambda \gg A_\kappa$.
The typical magnitude of the minimum is determined by $A_\lambda^4$ and the minimum becomes deeper as $\kappa$ becomes 
large and $\lambda$ becomes small.

%%%%%%%%%%%%%%%%%%%%%%%%%%%%%%%%%%%%%%%%%%%%%%%%%%%%%%%%%%%%%
\subsection{Unrealistic minimum along $H_2 \neq 0 $ direction}\label{sec:unre-minima-along-H2}
%%%%%%%%%%%%%%%%%%%%%%%%%%%%%%%%%%%%%%%%%%%%%%%%%%%%%%%%%%%%%
We consider the direction where only the up-type Higgs field is non-vanishing. 
In most cases, the up-type Higgs scalar has a tachyonic mass to achieve the EWSB in the NMSSM. Therefore there exists a 
minimum along which the up-type Higgs field develops a vev while other Higgs fields do not. 
On this minimum, the EWSB does not occur successfully because down-type fermions can not obtain masses.
Existence of this minimum was firstly studied in \cite{Ellwanger:1996gw}. 
Although this condition is quite important as we will show in the next section, it has not always 
been taken into account in the literature. 
Therefore, we reanalyze the unrealistic minimum along this direction and show an expression of a necessary 
condition for exclusion of parameters.

The scalar potential involving only the up-type Higgs field is given by
\begin{align}
V^{H_2} = m_{H_2}^2 |H_2|^2 + \frac{1}{8} g^2 |H_2|^4,\label{eq:7}
\end{align}
where $m_{H_2}^2$ is given by (\ref{eq:12}).
The extremal value of $|H_2|$ is obtained as
\begin{align}
 |H_2|_\mathrm{ext}^2 = - \frac{4 m_{H_2}^2}{g^2}, \label{eq:8}
\end{align}
and the minimum of the potential is given by
\begin{align}
 V^{ H_2}_\mathrm{min} = - \frac{2 (m_{H_2}^2)^2 }{g^2}. \label{eq:9}
\end{align}
If the minimum, (\ref{eq:9}), is deeper than the realistic minimum, 
the realistic minimum would not be realized.
Even if the realistic minimum might be realized once, 
it is unstable and may decay 
into this unrealistic minimum. Such a situation must be avoided by requiring 
\begin{align}
 V^{H_2}_{\mathrm{min}} \ge V_{\mathrm{min}}.\label{eq:10}
\end{align}

In the next section, we examine numerically the condition, (\ref{eq:10}).
Here, let us take a certain limit to reduce the number of free parameters 
appearing in the condition and to see intuitive implications 
of this condition, (\ref{eq:10}).
We may expect that $|m^2_{H_2}| \sim \mu^2 \gg m_Z^2$ in a typical 
parameter space.
The condition for parameters not satisfying the inequality, (\ref{eq:10}), can be expressed in a simple 
form when $\mu = \lambda s \gg m_Z$. The excluded region by the inequality (\ref{eq:10}) is given by
\begin{align}
 \kappa_- < \kappa < \kappa_+,\label{eq:14}
\end{align}
where
\begin{align}
 \kappa_\pm &= \frac{\lambda}{(g^2 - 2 \lambda^2 \cot^2 \beta) \mu} \times \bigg[ 
         -\frac{1}{6} g^2 A_\kappa - 2 \lambda^2 \mu' \cot\beta \nonumber \\
            &\quad \pm g \sqrt{ 2\lambda^2 \mu' (\mu' + \frac{1}{3} A_\kappa \cot\beta) + \frac{1}{36}g^2 A_\kappa^2}
       ~~\bigg], \label{eq:15}
\end{align}
with $\mu' = \mu - A_\lambda \cot\beta$. 
Furthermore, when $\mu \gg A_\lambda,~A_\kappa$ and $g \gg \lambda$, 
the value of $\kappa_+$ reduces to $\sqrt{2} \lambda^2/g$.
That is, the excluded region is approximately written as 
\begin{align}
 |\kappa| < \frac{\sqrt{2}}{g} \lambda^2, \label{eq:approx-H2>real}
\end{align}
which implies that a region with small $\kappa$ and large $\lambda$ is not allowed.
Alternatively, in the above limit, the depth of this minimum, (\ref{eq:9}), can be roughly estimated as 
\begin{align}
 V^{H_2}_{\mathrm{min}} \simeq -\frac{2}{g^2} \mu^4.
\end{align}
Then, comparing this with the realistic minimum, (\ref{eq:31}), 
the excluded region by (\ref{eq:10})  can be written as (\ref{eq:approx-H2>real}).

We have studied the unrealistic vacuum, where only 
$H_2$ develops its vev.
Similarly, we can study the unrealistic vacuum, where 
only the down-type Higgs field $H_1$ develops its vev, but 
the others, $H_2$ and $S$, have vanishing vevs.
The same results are applicable to (\ref{eq:7}), (\ref{eq:8})  and (\ref{eq:9})
except replacing $H_2$ and $m_{H_2}^2$ by $H_1$ and $m_{H_1}^2$.
That is, along this direction, 
the potential is written as 
\begin{align}
V^{H_1} = m_{H_1}^2 |H_1|^2 + \frac{1}{8} g^2 |H_1|^4,
\end{align}
and the extremal value of $H_1$ 
and the corresponding potential minimum 
would be obtained as 
\begin{align}
|H_1|_\mathrm{ext}^2 = - \frac{4 m_{H_1}^2}{g^2}, \qquad 
V^{ H_1}_\mathrm{min} = - \frac{2 (m_{H_1}^2)^2 }{g^2},
\end{align}
like (\ref{eq:8})  and (\ref{eq:9}) .
However, it is usually expected that $ m_{H_1}^2 > 0$.
In this case, such a minimum can not be realized.
Moreover, even if $ m_{H_1}^2 < 0$, it is expected that 
$ m_{H_1}^2 >  m_{H_2}^2$.
This implies that $V^{ H_1}_\mathrm{min} > V^{ H_2}_\mathrm{min}$.
Thus, the unrealistic minimum with $H_2 \neq 0$ is 
deeper and more important than one with $H_1 \neq 0$.

%%%%%%%%%%%%%%%%%%%%%%%%%%%%%%%%%%%%%%%%%%%%%%%%%%%%%%%%%%%%%
\subsection{Unrealistic minimum along $S \neq 0 $ direction}\label{sec:unre-minima-along-S}
%%%%%%%%%%%%%%%%%%%%%%%%%%%%%%%%%%%%%%%%%%%%%%%%%%%%%%%%%%%%%
We consider the direction along which only $S$ develops its vev. Along this direction, the sign of the 
trilinear term of $S$ can be taken as negative and therefore a minimum always exists. This minimum can be deeper than 
that of the realistic minimum.

The scalar potential along this direction is given by,
\begin{align}
 V^S(S) = \kappa^2 |S|^4 - \frac{2}{3}|\kappa| |A_\kappa| |S|^3 + m_S^2 |S|^2, \label{eq:17}
\end{align}
where $m_S^2$ is given by (\ref{eq:13}). Minimizing the potential, the extremal 
value of $S$ is obtained as 
\begin{align}
 |S|_{\mathrm{ext}} = \frac{|A_\kappa|}{4 |\kappa|} 
                     \left( 1 + \sqrt{1 - 8 \frac{m_S^2}{A_\kappa^2}} \right), \label{eq:18}
\end{align}
where $m_S^2 \le \frac{1}{8}A_\kappa^2$ should be satisfied.
Inserting (\ref{eq:18}) into (\ref{eq:17}), the minimum is given by
\begin{align}
 V^S_{\mathrm{min}} = -\frac{1}{6} |S|^2_{\mathrm{ext}}
                    \left( |\kappa||A_\kappa| |S|_{\mathrm{ext}} - 3 m_S^2 \right).
\end{align}
A necessary condition to avoid this minimum is to require
\begin{align}
 V^{S}_{\mathrm{min}} \ge V_{\mathrm{min}}. \label{eq:33}
\end{align}

In the next section, we examine numerically the condition, (\ref{eq:33}).
Here, let us see intuitive implications of this condition 
by using some approximation.
The depth of the minimum is roughly estimated as 
\begin{align}
 V^S_{\mathrm{min}} \simeq - \frac{A_\kappa^4}{384 \kappa^2}. \label{eq:32}
\end{align}
Comparing this minimum with (\ref{eq:31}), one can see that (\ref{eq:32}) can be deeper than the 
realistic minimum if
\begin{align}
 |\kappa| < \frac{5}{22} \frac{|A_\kappa|}{|\mu|} \lambda.
\end{align}

In the next section, we show that the conditions, (\ref{eq:10}) and (\ref{eq:33}) as well as requirement for 
 no tachyonic masses, exclude a large region of the parameter space.

%%%%%%%%%%%%%%%%%%%%%%%%%%%%%%%%%%%%%%%%%%%%%%%%%%%%%%%%%%%%%%%%%%%%%%%
\subsection{Other unrealistic and Charge and/or Colour Breaking minima}\label{sec:charge-andor-colour}
%%%%%%%%%%%%%%%%%%%%%%%%%%%%%%%%%%%%%%%%%%%%%%%%%%%%%%%%%%%%%%%%%%%%%%%
Finally, we present constraints from other unrealistic and CCB directions where not only squarks and/or 
sleptons but also the singlet scalar $S$ are non-vanishing. The directions we consider here are the same directions 
studied in the MSSM but are different by non-vanishing singlet scalar.
In the MSSM, CCB directions as well as UFB directions have been systematically studied in 
\cite{Frere:1983ag,AlvarezGaume:1983gj,Derendinger:1983bz,Kounnas:1983td,Claudson:1983et,Drees:1985ie,Gunion:1987qv,
Komatsu:1988mt,Gamberini:1989jw}
and general properties of these directions are summarized in \cite{Casas:1995pd}. One of the general properties of 
CCB and UFB directions is that $D$-terms, which are positive quartic terms, must be vanishing or kept under control. 
This property is important in particular 
when Yukawa couplings under consideration is smaller than the gauge coupling constants. 
In fact, the deepest CCB minima are found along vanishing $D$-term directions in the MSSM. 
Since the difference of the particle content of the NMSSM from that of the 
MSSM is the gauge singlet, vanishing $D$-term directions are the same as in the MSSM. Furthermore, when we consider 
the CCB and UFB directions of the MSSM with a vanishing singlet,
$S=0$, 
most of constraints to avoid such directions are obtained 
from those in the MSSM by setting $\mu=0$. Hence, non-trivial constraints are obtained along which the singlet scalar 
is non-vanishing, $S \neq 0$. For this purpose, it is enough to consider CCB and UFB directions of the MSSM with the singlet scalar. 
Indeed, the MSSM scalar potential has three UFB directions \cite{Casas:1995pd},
\begin{subequations}
\begin{align}
\mathrm{UFB-1:}&  \qquad |H_1| = |H_2| \neq 0, \\
\mathrm{UFB-2:}&  \qquad H_1,~H_2,~\tilde{L} \neq 0, \\
\mathrm{UFB-3:}&  \qquad H_2,~\tilde{L},~\tilde{Q},~\tilde{d}_R \neq 0, 
\quad \tilde{d}_L = \tilde{d}_R \equiv \tilde{d}, 
\end{align} 
\end{subequations}
where $\tilde{Q}$ and $\tilde{L}$ are chosen along $\tilde{d}_L$ and $\nuL$. 
Thus, we study each direction with adding non-vanishing gauge singlet, $S$ except for 
the UFB-1 direction that has already been studied in section \ref{sec:unre-minim-along-H1H2S}.
In addition, we study a typical CCB direction of the MSSM with 
adding non-vanishing gauge singlet.
In the following, we only present results along these directions.
Details of calculations are found in Appendix~\ref{sec:ccb-constr-nussm}.

The first direction is the one with the gauge singlet and the so-called 
MSSM UFB-2,
\begin{align}
  S,~H_1,~H_2,~\tilde{L} \neq 0, 
\end{align}
where $\tilde{L}$ is taken such that sneutrino is non-vanishing. A necessary condition is obtained by 
requiring that the minimum is positive,
\begin{align}
  \left(A_\lambda - \frac{1}{3} A_\kappa \right)^2 
\leq (1 + \gamma^2) \times 
\left[
m_{H_1}^2 - m_{\tilde{L}}^2 
+ m_S^2 \frac{\lambda}{\kappa}\frac{1}{\gamma}
+ (m_{H_2}^2 + m_{\tilde{L}}^2)\frac{1}{\gamma^2} 
\right],\label{eq:42}
\end{align}
where $\gamma$ is a real-positive parameter and smaller than $1$.
The most stringent constraint is obtained by minimizing the right-hand side. Then, the equation for the extremal 
value of $\gamma$ is given,
\begin{align}
 2 (m_{H_1}^2 - m_{\tilde{L}}^2) \gamma_\mathrm{ext}^4
+ m_S^2 \frac{\lambda}{\kappa}
\gamma_\mathrm{ext} (\gamma_\mathrm{ext}^2 - 1) 
- 2 (m_{H_2}^2 + m_{\tilde{L}}^2)
=0.
\end{align}

The second direction corresponds to the so-called MSSM UFB-3 direction 
with non-vanishing gauge singlet,
\begin{align}
  S,~H_2,~\tilde{L},~\tilde{Q},~\tilde{d}_R \neq 0, 
\quad \tilde{d}_L = \tilde{d}_R \equiv \tilde{d}, 
\end{align}
where $\tilde{Q}$ and $\tilde{L}$ are chosen along $\tilde{d}_L$ and $\nuL$. The vevs of $\tilde{d}_L$ and $\tilde{d}_R$ are chosen 
so that the $F$ term of $H_1$ vanishes. 
Then, following calculations in (\ref{sec:mssm-ufb-3}), we find a constraint 
\begin{align}
  |A_\kappa|^2
\leq 9 \Big[
\frac{(m_{\tilde{Q}}^2 + m_{\tilde{d}_R}^2 + m_{\tilde{L}}^2)^2}
{4 (m_{H_2}^2 + m_{\tilde{L}}^2)}
\frac{|\lambda|^2}{|Y_d|^2}
-2 (m_{H_2}^2 + m_{\tilde{L}}^2) + m_S^2 
\Big]. \label{eq:43}
\end{align}

The last direction we present is the so-called CCB-1 direction with the gauge singlet,
\begin{align}
  S,~H_2,~\tilde{Q},~\tilde{u}_R,~\tilde{L} \neq 0,
\end{align}
and a necessary condition obtained from this direction is 
\begin{align}
  \left(
\alpha^2 |A_u||Y_u| + \frac{1}{3}\sigma^3 |\kappa| |A_\kappa|
\right)^2 & \leq 
\left[
\sigma^2 (|\lambda|^2 + \sigma^2 |\kappa|^2)
+ |Y_u|^2 \alpha^2 (\alpha^2 + 2) 
\right] \nonumber \\
& \quad \times \left[
 m_{H_2}^2 + \alpha^2 
(m_{\tilde{Q}}^2 + m_{\tilde{u}_R}^2)
+ \sigma^2 m_S^2 + (1 - \alpha^2) \hat{m}_{\tilde{L}}^2
\right],\label{eq:35}
\end{align}
where $\alpha$ and $\sigma$ are real-positive numbers and $0 \le \alpha \le 1$.
The most stringent constraint is obtained by minimizing (\ref{eq:35}) with respect 
to $\alpha$ and $\sigma$, which can be done numerically.
Similarly, we can analyze other MSSM CCB directions with 
taking $S \neq 0$. At any rate, the constraints (\ref{eq:42}), (\ref{eq:43}) and (\ref{eq:35}) include more 
parameters such as $m_{\tilde{L}}^2$, $m_{\tilde{Q}}^2$, etc. 
Thus, we concentrate ourselves to analyze numerically 
constraints of section \ref{sec:unre-minim-along-H1H2S}, \ref{sec:unre-minima-along-H2} and \ref{sec:unre-minima-along-S} 
in the next section.

%%%%%%%%%%%%%%%%%%%%%%%%%%%%
\section{Numerical Analysis}\label{sec:numerical-analysis}
%%%%%%%%%%%%%%%%%%%%%%%%%%%%
We present numerical results of the constraints from the unrealistic minima, 
$V^{H_1H_2S}_{\mathrm{min}} \ge V_{\mathrm{min}}$ (\ref{eq:16}), 
$V^{H_2}_{\mathrm{min}} \ge V_{\mathrm{min}}$ (\ref{eq:10}) and  
$V^{S}_{\mathrm{min}} \ge V_{\mathrm{min}}$ (\ref{eq:33}) discussed in the previous section.
In addition to these constraints, we also take into
account the conditions that (i) 
physical masses of the neutral and the charged Higgs scalars are non-tachyonic, and (ii) the parameter, $\lambda$ and $\kappa$,
have no Landau poles until the GUT scale ($1.6 \times 10^{16}$ GeV).  For the condition (ii), 
we solve renormalization group (RG) equations at one-loop \cite{Derendinger:1983bz,Falck:1985aa} and require that $\lambda$ and $|\kappa|$ are 
smaller than $2\pi$ at the GUT scale \cite{Miller:2003ay,Ellwanger:2009dp}.
These constraints are usually studied in the literature, 
but not the constraints from (\ref{eq:16}), (\ref{eq:10}) and (\ref{eq:33}). 

\begin{figure}[t]
\begin{center}
\begin{tabular}{c}
 \includegraphics[height=60mm]{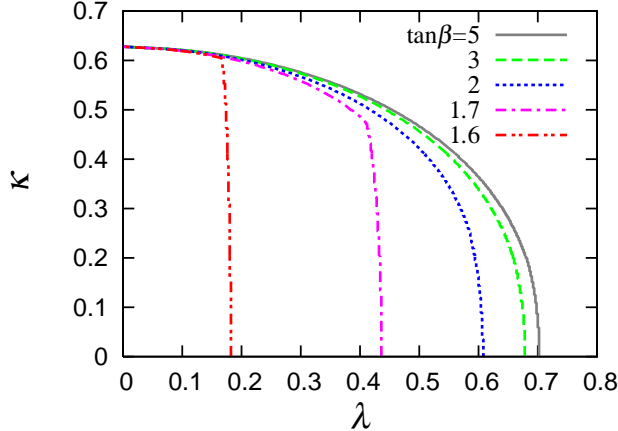}
\end{tabular} 
\caption{Region excluded by to the occurrence of a Landau pole on the $\lambda$-$\kappa$ plane. 
Solid (gray), dashed (green), dotted (blue) and dashed-dotted (violet), dashed-dotted-dotted (red) curves 
correspond to $\tan\beta=5,~3,~2$ and $1.7,~1.6$, respectively. The region outside each curve is excluded.}
\label{fig:landau-pole}
\end{center}
\end{figure}
Figure \ref{fig:landau-pole} shows the region excluded by the occurrence of a Landau pole on the $\lambda$-$\kappa$ plane. 
Solid (gray), dashed (green), dotted (blue) and dashed-dotted (violet), dashed-dotted-dotted (red) curves 
correspond to $\tan\beta=5,~3,~2$ and $1.7,~1.6$, respectively. The region outside each curve is excluded. 
One can see that $\lambda$ is more constrained as $\tan\beta$ is small, on the other hand the upper bound on 
$\kappa$ stays constant around $0.63$. This is because RG evolution of 
$\lambda$ is directly connected with the top Yukawa coupling. When $\tan\beta$ is small, 
the top Yukawa coupling at low energy is large and it grows quickly as the energy scale goes up. Then, $\lambda$ 
is driven to a large value as the top Yukawa coupling grows. On the other hand, RG evolution of $\kappa$ is proportional to 
$\kappa^2$ and depends on the top Yukawa coupling only through $\lambda$. Therefore, $\kappa$ starts to grow after 
the top Yukawa coupling and $\lambda$ becomes sufficiently larger than $2 \pi$. As we can see in figure \ref{fig:landau-pole}, 
the maximum value of $\lambda$ becomes small drastically for $\tan\beta < 2$ and it disappears when $\tan\beta \le 1.5$.
In the following, we choose moderate values of $\tan\beta$ to analyze the constraints from the unrealistic minima.

We use the parameter sets discussed in \cite{Cerdeno:2004xw,Cerdeno:2007sn} as illustrating examples.
In table \ref{tab:example-points}, we list four reference points taken from \cite{Cerdeno:2004xw,Cerdeno:2007sn} 
which lead to light spectra of Higgs scalars. Such a light Higgs scalar is characteristic in the NMSSM which is utilized 
for sneutrino dark matter \cite{Cerdeno:2008ep,Cerdeno:2009dv}. 
In fact, as we will see, the conditions, (\ref{eq:16}), (\ref{eq:10}) and (\ref{eq:33}), 
exclude larger regions on the $\lambda$-$\kappa$ plane.
\begin{table}[t]
\begin{center}
  \begin{tabular}{|c|c|c|c|c|} \hline
   ~~point~~~&~~~$\tan\beta$~~~&~~~$A_\lambda$~(GeV)~~~&~~~$A_\kappa$~(GeV)~~~&~~~$\mu=\lambda s$~(GeV)~~~ \\ \hline \hline
   ~~$1$~~& $3$ & $-200$ & $-50$ & $110$ \\ \hline
   ~~$2$~~& $3$ & $200$ & $-200$ & $110$ \\ \hline
   ~~$3$~~& $3$ & $50$ & $-50$ & $110$ \\ \hline
   ~~$4$~~& $5$ & $450$ & $50$ & $200$ \\ \hline
  \end{tabular}
 \caption{Reference points for numerical analysis taken from \cite{Cerdeno:2004xw,Cerdeno:2007sn}.}
\label{tab:example-points}
\end{center}
\end{table}

\begin{figure}[!t]
\begin{center}
\begin{tabular}{cc}
 \includegraphics[height=58mm]{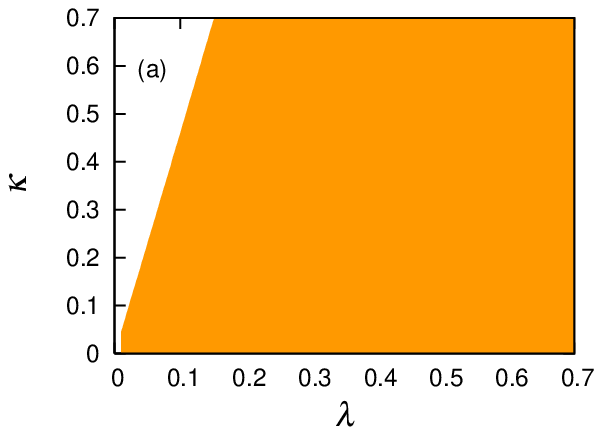} &
 \includegraphics[height=58mm]{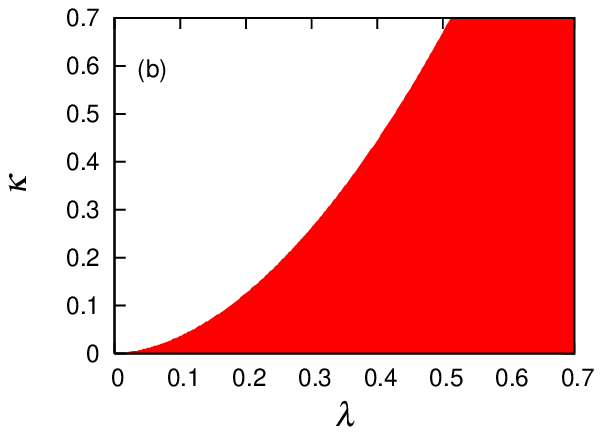} \\
 \includegraphics[height=58mm]{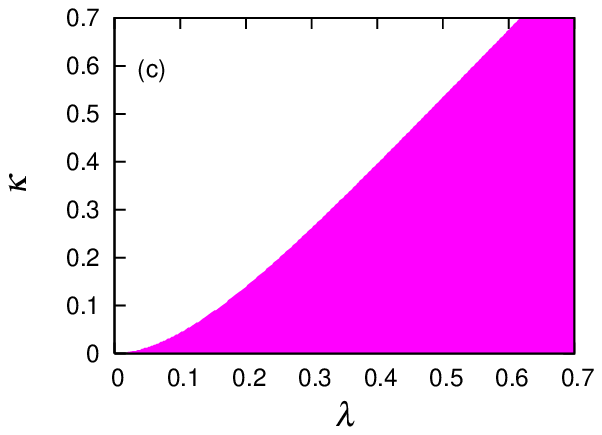} &
 \includegraphics[height=58mm]{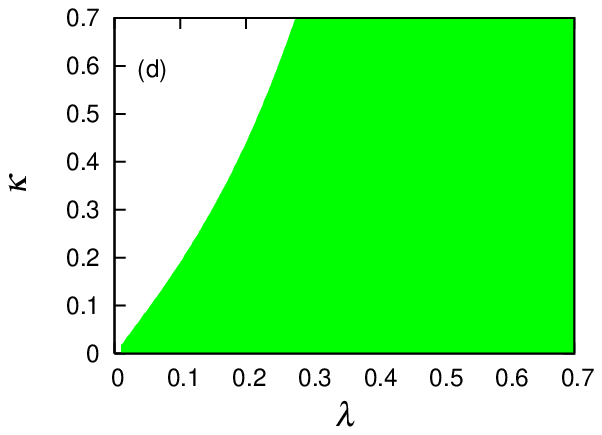} \\
 \multicolumn{2}{c}{\includegraphics[height=58mm]{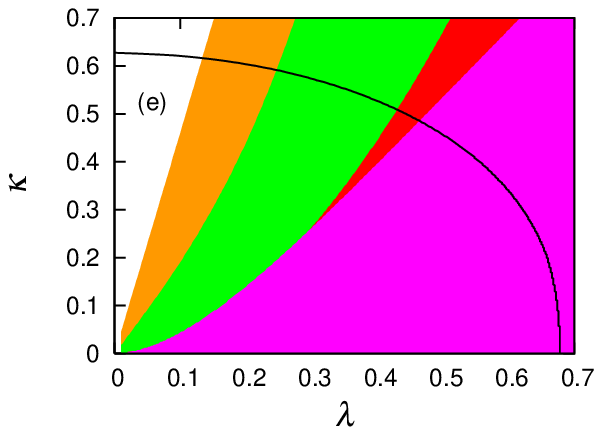}} 
\end{tabular} 
\caption{Excluded region for the point $1$. Region (coloured region) with $V^{H_1H_2S}_{\mathrm{min}} > V_{\mathrm{min}}$, 
$V^{H_2}_{\mathrm{min}} > V_{\mathrm{min}}$ and $V^{S}_{\mathrm{min}} > V_{\mathrm{min}}$ 
are shown in (a),
 (b) and (c), and the one with tachyonic Higgs masses is shown in (d). All conditions with Landau pole condition (black curve) 
are superposed in panel (e). Excluded region by the Landau pole is outside of the curve.}
\label{fig:point-1}
\end{center}
\end{figure}
Figure \ref{fig:point-1} shows regions excluded by $V^{H_1H_2S}_{\mathrm{min}} > V_{\mathrm{min}}$ (a), 
$V_{\mathrm{min}} > V_{\mathrm{min}}^{H_2}$ (b) and $V_{\mathrm{min}} > V_{\mathrm{min}}^{S}$ (c), tachyonic Higgs masses (d) 
on the $\lambda$-$\kappa$ plane.
In panel (e), all conditions in addition to the Landau pole condition are superposed.
In panel (a), we can see that the condition, $V^{H_1 H_2 S}_{\mathrm{min}} > V_{\mathrm{min}}$, excludes a wider region 
for a large value of $\lambda$. One may consider that this minimum can be a realistic minimum because three Higgs fields develop vevs. 
However, as we shown in figure \ref{fig:landau-pole}, no region is allowed for $\tan\beta=1$ by an occurrence of a Landau pole.
The reason why a region of large $\lambda$ is excluded is as follows. From (\ref{eq:37}), $V^{H_1 H_2 S}_{\mathrm{min}}$ becomes shallower 
as $\lambda$ becomes larger and its depth is typically determined by $A_\lambda^4$. However, the realistic minimum has a similar 
dependence and its depth is determined by $A_\kappa \mu^3$ as shown in (\ref{eq:31}). Therefore, 
$V^{H_1 H_2 S}_{\mathrm{min}}$ becomes deeper than the realistic minimum for $|A_\lambda| > A_\kappa,~\mu$. Furthermore, (\ref{eq:40}) is 
negative in a wide region of parameter space when $A_\lambda$ is negative, and the minimum, $V^{H_1 H_2 S}_{\mathrm{min}}$ (\ref{eq:27}), 
appears in that region. Thus, this condition can exclude a large region of the parameter space when $A_\lambda$ is 
large and negative. 

Panel (b) shows that the condition, $V^{H_2}_{\mathrm{min}} > V_{\mathrm{min}}$, also excludes 
a wider region for large $\lambda$. This can be understood as follows; recall from (\ref{eq:9}) and (\ref{eq:12}) 
that $V^{H_2}_{\mathrm{min}} \propto -(m_{H_2}^2)^2$ and $|m_{H_2}^2| \sim \lambda^2 m_Z^2$. Therefore $V^{H_2}_{\mathrm{min}}$ 
becomes deeper as $\lambda$ increases. 
On the other hand, from (\ref{eq:31}), $V_{\mathrm{min}} \simeq \overline{V}^S_{\mathrm{min}}$ is a polynomial of $\mu/\lambda$.
For a fixed value of $\mu$, the value $|\overline{V}^S_{\mathrm{min}}|$ decreases as $\lambda$ increases.
Hence, the minimum value of $\overline{V}^S_{\mathrm{min}}$ as well as $V_{\mathrm{min}}$  becomes 
shallower as $\lambda$ increases.
In addition, the $\lambda$ independent term of $V^{H_2}_\mathrm{min}$
is $-\mu^4/(g_1^2 + g_2^2)$ while that of $\overline{V}^S_{\mathrm{min}}$ appears in 
$m_S^2 s^2$, i.e. 
$A_\lambda \mu m_Z^2/(g_1^2 + g_2^2) \sin 2\beta$. 
It is expected that $-\mu^4/(g_1^2 + g_2^2) < A_\lambda \mu m_Z^2/(g_1^2 + g_2^2) \sin 2\beta$. 
Thus, $V^{H_2}_{\mathrm{min}}$ is deeper than $V_{\mathrm{min}}$ 
for a large value of $\lambda$. This result can be seen more easily in (\ref{eq:15}). When $\lambda$ is large, the approximate 
condition, $\kappa_+$, becomes
\begin{align}
 \kappa_+ \simeq \frac{\lambda^2}{g^2 \mu}
            ( \sqrt{2} g |\mu'| -2 \lambda \mu'\cot\beta),
\end{align}
where we used $\mu' \gg g^2 A_\kappa$.

In panel (c), the region excluded by $V^S_{\mathrm{min}} > V_{\mathrm{min}}$ is shown. It is seen 
that this condition also excludes a larger region for large
$\lambda$. The reason is almost the same as for the panel (b).
The value $|\overline{V}^S_{\mathrm{min}}|$ in $V_{\mathrm{min}}$ decreases as $\lambda$
increases. 
On the other hand, the $\lambda$ dependence of $V^S_{\mathrm{min}}$ 
appears through $m_S^2$, which also decreases as $\lambda$ increases.
Smaller value of $m_S^2$ increases $|V^S_{\mathrm{min}}|$.
Thus, the region with large  $\lambda$ is excluded by 
the constraint $V^S_{\mathrm{min}} > V_{\mathrm{min}}$.

The region excluded by tachyonic Higgs masses is shown in panel (d). In this region, deeper minima than the realistic one 
exist and hence the EWSB is unstable. Such minima will be $V^{H_1 H_2 S}_{\mathrm{min}}$, $V^{H_2}_{\mathrm{min}}$ and $V^{S}_{\mathrm{min}}$.
The reason for tachyonic masses is simply due to large off-diagonal elements in the mass matrices. From the mass matrices, 
(\ref{eq:23}) and (\ref{eq:24}), one can understand that the off-diagonal elements become larger than or comparable to the 
diagonal elements when $\lambda$ is large, which leads to tachyonic masses after diagonalization of the mass matrices. 
Thus, again, large values of $\lambda$ are excluded. 

In the end, in panel (e), we superpose all constraints with that 
for avoiding an occurrence of Landau poles. 
The region excluded by Landau poles of $\lambda$, $\kappa$ and the top Yukawa coupling 
is indicated by outside the solid (black) curve.
One can see that large values of $\lambda$ are excluded by unphysical minima and 
tachyonic masses while large values of $\kappa$ are excluded by the
Landau pole condition. 
One can also see that the condition,  
$V^{H_1 H_2 S}_{\mathrm{min}} > V_{\mathrm{min}}$, 
is the tighter constraint than the one from the tachyonic condition. This means that even if the EWSB vacuum has no tachyonic 
directions, deeper minimum along $H_1 = H_2 \neq 0$ and $F_S = 0$ direction exists and makes the EWSB vacuum unstable. 
Therefore it is important to take into account this constraint for the analysis. In the end, it is worthwhile to mention 
that the allowed region for this point is located within $\lambda \le 0.15$ and $\kappa \le 0.62$.

\begin{figure}[t]
\begin{center}
\begin{tabular}{cc}
 \includegraphics[height=58mm]{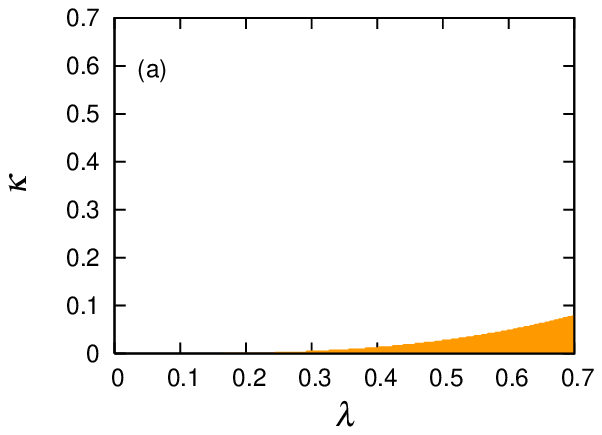} &
 \includegraphics[height=58mm]{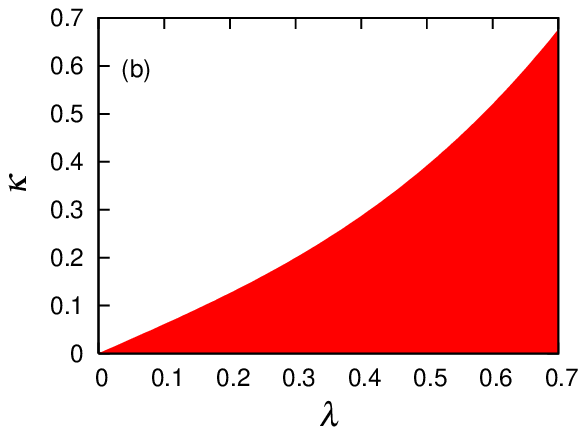} \\
 \includegraphics[height=58mm]{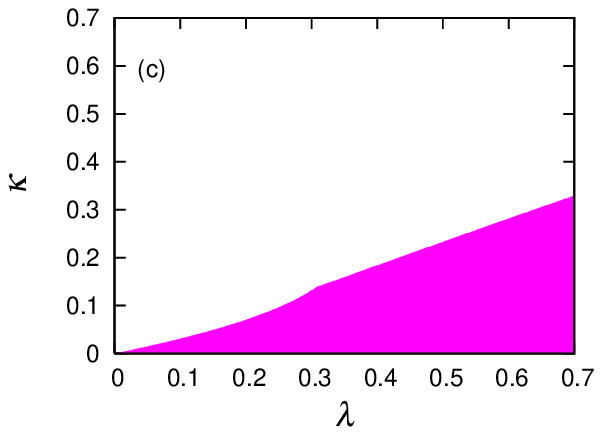} &
 \includegraphics[height=58mm]{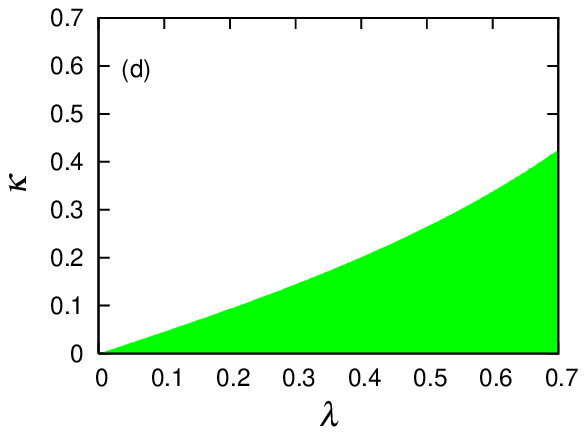} \\
 \multicolumn{2}{c}{\includegraphics[height=58mm]{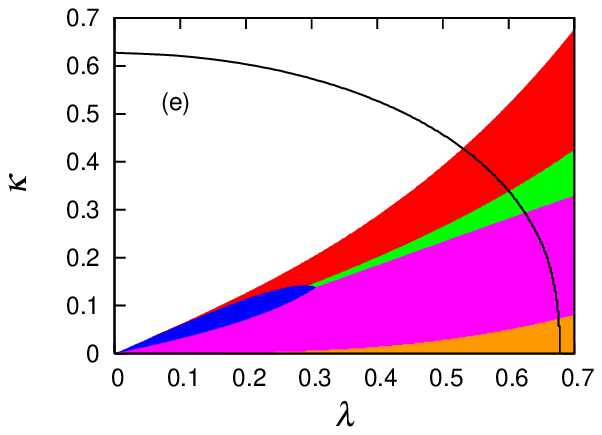}}
\end{tabular} 
\caption{Excluded region for the point $2$. Each panels are the same as the point $1$. 
In panel (e), a blue region indicates excluded region that the value $V_{\mathrm{min}}$ is positive.}
\label{fig:point-2}
\end{center}
\end{figure}
Figure \ref{fig:point-2} shows the excluded region for the point $2$. Each panel corresponds to the same constraints 
of the unrealistic minima as in figure \ref{fig:point-1}. In panel (a), we see that the condition, 
$V^{H_1 H_2 S}_{\mathrm{min}} > V_{\mathrm{min}}$, is not as tight as that for the point $1$. As explained in figure \ref{fig:point-1}, 
$\hat{m}^2$ in (\ref{eq:40}) is positive in a large region of the parameter space for a positive $A_\lambda$, and hence 
$V^{H_1 H_2 S}_{\mathrm{min}}$ does not appear.
In panel (c), we also see that the region excluded by $V^S_{\mathrm{min}} > V_{\mathrm{min}}$ is weaker than that for the point $1$. 
This is because $m_S^2$ is positive and larger than $\frac{1}{8}A_\kappa^2$ in a wide region of the parameter space, and hence 
$V^{S}_{\mathrm{min}}$ does not appear. Panel (d) shows all constraints with that for avoiding an occurrence of Landau poles.
In addition, the blue region corresponds to the region, where $V_{\mathrm{min}}$ is positive. 
We can see that the condition, $V^{H_2}_{\mathrm{min}} > V_{\mathrm{min}}$, is the tighter constraint than the one from the 
tachyonic condition. This means that deeper minimum appears along $H_2 \neq 0$ direction and the EWSB vacuum becomes unstable. 
Therefore the constraint, (\ref{eq:10}), is also important for the analysis. Again, the allowed region for this point is 
located within $\lambda \le 0.5$ and $\kappa \le 0.62$.

\begin{figure}[t]
\begin{center}
\begin{tabular}{cc}
 \includegraphics[height=58mm]{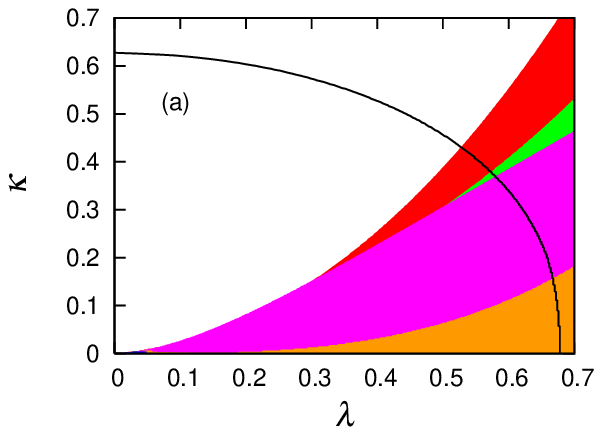} & 
 \includegraphics[height=58mm]{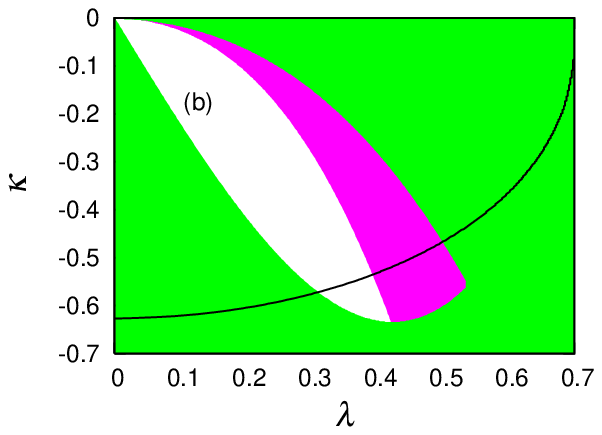} \\
\end{tabular} 
\caption{Excluded regions for the points $3$ and $4$. Panel (a) is for the point $3$ and (b) is for $4$. Colour indicates 
the same condition of the panels of Fig.~\ref{fig:point-2}.}
\label{fig:point-3-4}
\end{center}
\end{figure}
In Figure \ref{fig:point-3-4}, the same figures of Fig.~\ref{fig:point-1}.(e) are shown for the points $3$ and $4$. In panel (a), 
it is seen that the difference from the point $2$ is that a wider region for large $\lambda$ is excluded by the conditions of 
the unrealistic minima and the tachyonic masses. This result is non-trivial and due to complicated dependences on the parameters. 
The trilinear couplings $A_\lambda$ and $A_\kappa$ for the point $3$ are smaller than those for the point $2$. The smaller 
couplings result in shallow depths of not only the unrealistic minima but also the realistic minimum. The realistic minimum tends 
to be shallower than the unrealistic ones do, for smaller trilinear couplings.
In panel (b), a region for negative $\kappa$ is largely excluded because $A_\kappa$ is negative. It is seen that a wide region of large 
$\lambda$ is excluded by the condition, $V^{S}_{\mathrm{min}} > V_{\mathrm{min}}$, and that for small $\lambda$ is excluded by 
the condition from tachyonic masses. Thus, the condition, $V^{S}_{\mathrm{min}} > V_{\mathrm{min}}$, is also important. 

%\begin{figure}
%\begin{center}
%\begin{tabular}{cc}
% \includegraphics[height=58mm]{figures/appro-kappa/appro-kappa.tanb=3.Al=200.Ak=-200.mu=110.eps} & 
% \includegraphics[height=58mm]{figures/appro-kappa/appro-kappa.tanb=3.Al=200.Ak=-200.mu=400.eps} \\
%\end{tabular} 
%\caption{Comparison of the approximate formula, (\ref{eq:14}) with full numerical analysis. In panel (a) and (b), 
%$\mu$ is taken as $110$ and $400$ GeV, respectively.}
%\label{fig:appro-kappa}
%\end{center}
%\end{figure}
%In Figure \ref{fig:appro-kappa}, the approximate formula for the $H_2$ direction, (\ref{eq:14}), is 
%shown with the results of full analysis. 
%The effective supersymmetric Higgs mass, $\mu$, is taken as $110$ GeV
%for the panel (a) and $400$ GeV for the panel (b). Other 
%parameters are taken as $\tan\beta=3$, $A_\lambda=200$ GeV and 
%$A_\kappa = -200$ GeV. Red (solid) curve represents the approximate 
%formula and dashed (green) curve represents the result of full
%analysis. 
%The parameter regions under these curves correspond to 
%the excluded regions.
%In panel (a), one can see that the approximate 
%formula is in good agreement for $\lambda \le 0.2$ and shows a significant deviation for $\lambda \ge 0.4$. In panel (b), 
%the approximate formula shows good agreement with the full analysis for all region of $\lambda$. The approximate formula 
%is especially useful for the Constrained NMSSM because $\mu$ usually becomes larger than $m_Z$ due to a large tachyonic mass of the 
%up-type Higgs scalar.

%%%%%%%%%%%%%%%%%%%%%%%%%%%%%%%%
\section{Conclusion and Discussion}\label{sec:summary-discussion}
%%%%%%%%%%%%%%%%%%%%%%%%%%%%%%%%
In this paper, we have analyzed scalar potentials of the NMSSM and 
systematically studied constraints from unphysical minima on which the EWSB does not 
occur successfully and CCB minima on which colour and/or electric charge symmetry is broken. 

In section \ref{sec:constr-from-vacua}, we derived explicit expressions of the 
unrealistic minima along which $H_1 = H_2 \neq 0$ with $F_S = 0$ and only one of the three Higgs 
fields develop their vevs. These unrealistic minima threaten the realization of the EWSB if these are deeper than 
the EWSB vacuum. Indeed, using some approximation, we showed such minima can be deeper than the realistic minima 
for a large value of $\lambda$. Constraints from other directions involving squarks and sleptons are also derived.

In section \ref{sec:numerical-analysis}, we numerically studied the constraints to avoid the unrealistic minima as 
well as an occurrence of Landau poles and tachyonic masses. For the condition to avoid Landau poles, it was found 
in figure \ref{fig:landau-pole} that the parameter, $\lambda$, is significantly excluded for $\tan\beta < 2$ and the 
allowed region disappears for $\tan\beta < 1.5$. Thus, the direction with $H_1 = H_2 \neq 0$ can never be the realistic 
minimum under the Landau pole condition.
Then, we chose four points of the parameters with moderate value of $\tan\beta$ as illustrating examples, 
which correspond to the light spectra of Higgs scalars. In figure \ref{fig:point-1}, we showed that most of region on the 
$\lambda$-$\kappa$ plane are ruled out for the point $1$. The most stringent constraints for this point come from 
$H_1 = H_2 \neq 0$ with $F_S = 0$ direction and the absence of Landau poles. This result is rather general for points with 
large and negative $A_\lambda$ because this unrealistic minimum widely appears for the negative $A_\lambda$ and 
its depth is determined by $A_\lambda^4$. 
In figure \ref{fig:point-2}, it was shown that the constraints from $H_2$ direction and Landau poles 
exclude a wider region of large $\lambda$ and $\kappa$ in the point $2$. This is because $V^{H_2}_\mathrm{min}$ becomes 
deeper while $V_{\mathrm{min}}$ becomes shallower as $\lambda$ increases. 
Similar result is obtained in figure \ref{fig:point-3-4}.(a) for the point $3$. 
In figure \ref{fig:point-3-4}.(b), we also found that a large region of the parameter space is ruled out for negative $\kappa$. 
The stringent constraints are the ones from $S$ direction and tachyonic masses. 
Our results imply that each of the constraints is significant. Since the constraints are independent of each other, 
it is important to apply all constraints considered here in the analyses of the NMSSM.

In the end, we discuss an implication of our results. The coupling constant $\lambda$ is important to increase the Higgs mass at the
tree-level. However, from our numerical analysis, we saw that a large value of $\lambda$ is not allowed from the constraints. 
This implies that the SM-like Higgs mass would be similar to that of the MSSM at the tree-level. 
Thus, our constraints would be quite important to the spectrum of the Higgs sector as well as dark matter physics.
We would study these aspects elsewhere.

\acknowledgments
T.~K. is supported in part by the Grant-in-Aid for Scientific Research No.~20540266 and the Grant-in-Aid for the 
Global COE Program ``The Next Generation of Physics, Spun from Universality and Emergence'' from the Ministry of 
Education, Culture, Sports, Science and Technology of Japan. 
O.~S. is supported in part by the scientific research grants from Hokkai-Gakuen and thanks the Yukawa Institute for 
Theoretical Physics at Kyoto University, where this work was completed during the YITP-T-10-05 on ``Cosmological Perturbation 
and Cosmic Microwave Background''.
T.~S. is the Yukawa Fellow and the work of T.~S. is partially supported 
by Yukawa Memorial Foundation. The work of Y.~Kanehata and Y.~Konishi is supported by Maskawa Institute at Kyoto Sangyo 
University.

\appendix
%%%%%%%%%%%%%%%%%%%%%%%%%%
\section{Scalar potential}\label{sec:scalar-potential}
%%%%%%%%%%%%%%%%%%%%%%%%%%
In this Appendix, we give notations of scalars and the scalar potential
of the NMSSM. Throughout this article, flavour Indices are suppressed for simplicity.

The down-type and the up-type Higgs scalars are denoted as 
\begin{align}
 H_1 =
  \begin{pmatrix}
   H^1_1 \\
   H^2_1
  \end{pmatrix},\quad 
 H_2 = 
  \begin{pmatrix}
   H^1_2 \\
   H^2_2
  \end{pmatrix},
\end{align}
where $H_1^1,~H_2^2$ are electrically neutral components and $H_1^2$($H_2^1$) are negatively(positively) charged 
components. The gauge singlet scalar is denoted by $S$. The left-handed and the right-handed squarks are  
denoted as
\begin{align}
 \tilde{Q} =
  \begin{pmatrix}
   \tilde{u}_L \\
   \tilde{d}_L
  \end{pmatrix},\quad
 \tilde{u}_R,~\tilde{d}_R,
\end{align}
and those of sleptons are denoted as 
\begin{align}
 \tilde{L} =
  \begin{pmatrix}
   \tilde{\nu}_L \\
   \tilde{e}_L
  \end{pmatrix},\quad
 \tilde{e}_R.
\end{align}

The superpotential of the NMSSM is given by 
\begin{align}
 \mathcal{W}_{\mathrm{NMSSM}} &= Y_d \hat{H}_1\cdot \hat{Q} \hat{D}^c_R 
                              + Y_u \hat{H}_2\cdot \hat{Q} \hat{U}^c_R 
                              + Y_e \hat{H}_1\cdot \hat{L} \hat{E}^c_R
                              - \lambda\hat{S} \hat{H}_1\cdot\hat{H}_2 
                              + \frac{1}{3}\kappa\hat{S}^3,
\end{align}
where a symbol ``hat'' denotes a superfield of each chiral multiplet and a symbol ``dot'' represents an 
inner product for $SU(2)$ doublets, $A \cdot B \equiv A^1B^2 - A^2 B^1$.  The Yukawa coupling constants for up-type quarks, 
down-type quarks and leptons are denoted by $Y_u$, $Y_d$ and $Y_e$, respectively, and that for the singlet fermion is $\lambda$. 
The self-coupling constant of the singlet is $\kappa$.
The scalar potential, $V$, is divided into three parts which consist
of $F$, $D$ and the soft SUSY breaking terms,
\begin{align}
 V = V_F + V_D + V_{\mathrm{soft}}.
\end{align}
The $F$ term potential, $V_F$, is given by a sum of absolute square of all matter auxiliary fields,
\begin{align}
 V_F = \sum_{i=\mathrm{matter}} |F_i|^2,
\end{align}
where
\begin{subequations}
\begin{align}
 F_{H_1^1}^\ast&=
 - \lambda s H_2^2 + Y_{d} {\tilde d}_{L} {\tilde d}_{R}^\ast + Y_{e} {\tilde e}_{L} {\tilde e}_{R}^\ast, \\
 F_{H_1^2}^\ast&=
 \lambda s H_2^1 - Y_{d} {\tilde u}_{L}{\tilde d}_{R}^\ast -Y_{e} {\tilde \nu}_{L} {\tilde e}_{R}^\ast, \\
 F_{H_2^1}^\ast&=
 \lambda s H_1^2 + Y_{u} {\tilde d}_{L} {\tilde u}_{R}^\ast, \\
 F_{H_2^2}^\ast&=
 - \lambda s H_1^1 - Y_{u} {\tilde u}_{L} {\tilde u}_{R}^\ast, \\
 F_{S}^\ast &=
  \lambda \left(H_1^1 H_2^2 - H_1^2 H_2^1\right) - \kappa s^2, \\
 F_{{\tilde d}_{L}}^\ast&=
 Y_{d} H_1^1 {\tilde d}_{R}^\ast + Y_{u} H_2^1 {\tilde u}_{R}^\ast, \\
 F_{{\tilde u}_{L}}^\ast&=
 - Y_{d} H_1^2 {\tilde d}_{R}^\ast - Y_{u} H_2^2 {\tilde u}_{R}^\ast, \\
 F_{{\tilde e}_{L}}^\ast&=
 Y_{e} H_1^1 {\tilde e}_{R}^\ast, \\
 F_{{\tilde \nu}_{L}}^\ast&=
 - Y_{e} H_1^2 {\tilde e}_{R}^\ast, \\
 F_{{\tilde d}_{R}}&=
 Y_{d} \left(H_1^1 {\tilde d}_{L} - H_1^2 {\tilde u}_{L}\right), \\
 F_{{\tilde u}_{R}}&=
 Y_{u} \left(H_2^1 {\tilde d}_{L} - H_2^2 {\tilde u}_{L}\right), \\
 F_{{\tilde e}_{R}}&=
 Y_{e} \left(H_1^1 {\tilde e}_{L} - H_1^2 {\tilde \nu}_{L}\right).
\end{align}
\end{subequations}
The $D$ term potential, $V_D$, is given by a sum of square of all gauge auxiliary fields,
\begin{align}
 V_D = \frac{1}{2} \bigg( (D^a_{SU(3)})^2 + (D^a_{SU(2)})^2 + (D_{U(1)})^2 \bigg),
\end{align}
where the subscripts represent gauge groups and $a$ runs from $1$ to $8~(3)$ for $SU(3)~(SU(2))$. 
Summation over $a$ should be understood. The auxiliary fields are given by  
\begin{subequations}
\begin{align}
 D^a_{SU(3)} &= g_3 \left( \tilde{Q}^\dagger \frac{\lambda^a}{2} \tilde{Q} 
                          - \tilde{u}^\ast_R \frac{\lambda^a}{2} \tilde{u}_R
                          - \tilde{d}^\ast_R \frac{\lambda^a}{2} \tilde{d}_R \right), \\
 D^a_{SU(2)} &= g_2 \big( \tilde{Q}^\dagger T^a \tilde{Q} + \tilde{L}^\dagger T^a \tilde{L}
                + H_1^\dagger T^a H_1 + H_2^\dagger T^a H_2 \big),\\
 D_{U(1)} &= g_1 \left( \frac{1}{6} \tilde{Q}^\dagger \tilde{Q} -\frac{2}{3} \tilde{u}_R^\ast \tilde{u}_R 
           + \frac{1}{3} \tilde{d}^\ast_R \tilde{d}_R -\frac{1}{2}\tilde{L}^\dagger \tilde{L}
           + \tilde{e}_R^\ast \tilde{e}_R -\frac{1}{2}H_1^\dagger H_1 + \frac{1}{2}H_2^\dagger H_2 \right),
\end{align}
\end{subequations}
where $g_i~(i=1,2,3)$ are gauge coupling constants, and $\lambda^a$ and $T^a$ are Gell-Mann and Pauli matrices, respectively.
The scalar potential, $V_{\mathrm{soft}}$, with soft SUSY breaking
terms is given as
\begin{align}
  V_{soft} & = m^2_{H_1} H_1^\dagger H_1 + m^2_{H_2} H_2^\dagger H_2 + m^2_{S} S^\dagger S 
              + \left(\frac{1}{3} \kappa A_\kappa S^3 - \lambda
              A_\lambda S H_1 \cdot H_2 + h.c.\right) \nonumber \\
           &~~+ m^2_{\tilde{Q}} \tilde{Q}^\dagger \tilde{Q} + m^2_{\tilde{u}_R} \tilde{u}_R^\ast \tilde{u}_R
              + m^2_{\tilde{d}_R} \tilde{d}_R^\ast \tilde{d}_R 
              + m^2_{\tilde{L}} \tilde{L}^\dagger \tilde{L} + m^2_{\tilde{e}_R} \tilde{e}_R^\ast \tilde{e}_R  \nonumber \\
           &~~+ \big(A_d Y_d H_1 \cdot \tilde{Q} \tilde{d}_R^\ast + A_u Y_u H_2 \cdot \tilde{Q} \tilde{u}_R^\ast
              + A_e Y_e H_1 \cdot \tilde{L} \tilde{e}_R^\ast + h.c.\big), 
\end{align}
where $A_\lambda$ and $A_\kappa$ are trilinear couplings for Higgs fields 
and $A_i~(i=u,d,e)$ are those for squarks and sleptons. The soft SUSY breaking masses squared are denoted by 
$m^2_i~(i=H_1, H_2, \tilde Q, \uR, \dR, \tilde L, \tilde e_R)$.

%%%%%%%%%%%%%%%%%%%%%%%%%%%%%%%%%%%%%%%%%%%%%%%%%%%%%%%%%%%%%%%%%%%%%%%%%%%
\section{Constraints from Charge and/or Colour Breaking Minima in the NMSSM}\label{sec:ccb-constr-nussm}
%%%%%%%%%%%%%%%%%%%%%%%%%%%%%%%%%%%%%%%%%%%%%%%%%%%%%%%%%%%%%%%%%%%%%%%%%%
In this Appendix, we derive constraints to avoid CCB minima discussed in the Sec.~\ref{sec:charge-andor-colour}.
In the following, we note $H_1^1$ and $H_2^2$ as $H_1$ and $H_2$ for abbreviation.

%%%%%%%%%%%%%%%%%%%%%%%%%%%%%%%%%%%%%%%%%%%%%%%%%%%%%%%%%
\subsection{MSSM UFB-2 direction with gauge singlet}\label{sec:mssm-ufb-2}
%%%%%%%%%%%%%%%%%%%%%%%%%%%%%%%%%%%%%%%%%%%%%%%%%%%%%%%%%
We analyze the so-called MSSM UFB-2 direction with the gauge singlet, 
\begin{align}
 S,~H_1,~H_2,~\tilde{L} \neq 0, 
\end{align}
where $\tilde{L}$ is chosen along $\nuL$. 
The scalar potential along this direction is 
\begin{align}
V_\mathrm{UFB-2} &= \lambda^2 |S|^2 (|H_1|^2 + |H_2|^2) + |F_S|^2 
                   + m_{H_1}^2 |H_1|^2 + m_{H_2}^2 |H_2|^2 + m_{\tilde{L}}^2 |\tilde{\nu}_L|^2 + m_S^2 |S|^2 \nonumber \\
                 &\quad - 2 \lambda A_\lambda H_1 H_2 S  - \frac{2}{3} \kappa A_\kappa S ^3 
                   + \frac{1}{8}g^2(|H_2|^2 - |H_1|^2 - |\tilde{\nu}_L|^2)^2,
\end{align}
where $F_S$ is given by (\ref{eq:28}). By minimizing $|\tilde{\nu}_L|$, we have
\begin{align}
 |\tilde{\nu}_L| = -\left( 4 \frac{m_{\tilde{L}}^2}{g^2} - |H_2|^2 + |H_1|^2  \right),
\end{align}
where $4 m_{\tilde{L}}^2/g^2 - |H_2|^2 + |H_1|^2 < 0$ must be satisfied.
Inserting this into the potential, the potential is given by
\begin{align}
 V_{\mathrm{UFB-2}} &= -2 \frac{m_{\tilde{L}}^4}{g^2} + \lambda^2 |S|^2 ( |H_1|^2 + |H_2|^2 ) + |F_S|^2 \nonumber \\
                    &\quad + ( m_{H_1}^2 - m_{\tilde{L}}^2 ) |H_1|^2 + ( m_{H_2}^2 + m_{\tilde{L}}^2 ) |H_2|^2 + m_S^2 |S|^2 \nonumber \\ 
                    &\quad - 2 \lambda A_\lambda S H_1 H_2 + \frac{2}{3} \kappa A_\kappa S^3.
\end{align}
Choosing $S$ so that $F_S$ is vanishing as (\ref{eq:41}) and parameterizing the 
vev as $|H_1| = \gamma |H_2|$ then the potential is written in a form similar to (\ref{eq:16}),
\begin{align}
 V_{\mathrm{UFB-2}} = \hat{F}|H_2|^4 -2 \hat{A} |H_2|^3 + \hat{m}^2 |H_2|^2 - 2 \frac{m_{\tilde{L}}^4}{g^2},
\end{align}
where
\begin{subequations}\label{eq:29}
\begin{align}
 \hat{F}  &= \frac{\lambda^3}{\kappa} \gamma (1 + \gamma^2), \\
 \hat{A}  &= \left(|A_\lambda - \frac{1}{3} A_\kappa|\right) 
                    \sqrt{\frac{ \lambda ^3}{ \kappa }\gamma^3}, \\
\hat{m}^2 &= ( m_{H_1}^2 - m_{\tilde{L}}^2 ) \gamma^2 + ( m_{H_2}^2 + m_{\tilde{L}}^2 ) + m_S^2 \frac{\lambda}{\kappa}\gamma.
\end{align}
\end{subequations}
When the constant term in the potential is negligible, the extremal value of $|H_2|$ and the depth of the minimum are obtained by simply replacing 
$\hat{F}$, $\hat{A}$ and $\hat{m}^2$ in (\ref{eq:26}) and (\ref{eq:27}) with (\ref{eq:29}).
The typical magnitudes of the extremal value and the depth of the minimum are similar to the one in \ref{sec:unre-minim-along-H1H2S}.
To avoid the minimum, we obtain a constraint
\begin{align}
 \left(A_\lambda - \frac{1}{3} A_\kappa \right)^2 
\leq (1 + \gamma^2) \times 
\left[
m_{H_1}^2 - m_{\tilde{L}}^2 
+ m_S^2 \frac{\lambda}{\kappa}\frac{1}{\gamma}
+ (m_{H_2}^2 + m_{\tilde{L}}^2)\frac{1}{\gamma^2} 
\right] .
\end{align}

Minimizing the right hand side with respect to $\gamma$, 
we find that the extremal value of the $\gamma$ has to satisfy the
following relation:
\begin{align}
2 (m_{H_1}^2 - m_{\tilde{L}}^2) \gamma_\mathrm{ext}^4
+ m_S^2 \frac{\lambda}{\kappa}
\gamma_\mathrm{ext} (\gamma_\mathrm{ext}^2 - 1) 
- 2 (m_{H_2}^2 + m_{\tilde{L}}^2)
=0.
\end{align}

%%%%%%%%%%%%%%%%%%%%%%%%%%%%%%%%%%%%%%%%%%%%%%%%%%%%%
\subsection{MSSM UFB-3 direction with gauge singlet }\label{sec:mssm-ufb-3}
%%%%%%%%%%%%%%%%%%%%%%%%%%%%%%%%%%%%%%%%%%%%%%%%%%%%%
Here, we study the so-called MSSM UFB-3 direction with 
the gauge singlet $S$. 
That is, the direction we analyze is 
\begin{align}
 S,~H_2,~\tilde{L},~\tilde{Q},~\tilde{d}_R \neq 0, 
\quad \tilde{d}_L = \tilde{d}_R \equiv \tilde{d}, 
\end{align}
where $\tilde{Q}$ and $\tilde{L}$ are chosen along $\tilde{d}_L$ and $\nuL$. The vevs of $\tilde{d}_L$ and $\tilde{d}_R$ are chosen 
so that the $F$ term of $H_1$ vanishes. 
Using the parametrization,
\begin{align}
 |\tilde{L}| = \gamma_L |H_2|, \quad
 |S| = \sigma |H_2|,
\end{align}
the condition for $F_{H_1}=0$ is expressed as,
\begin{eqnarray}
 \gamma_L^2 = 1 + \frac{\lambda}{|Y_d|}\sigma.
\end{eqnarray}
The scalar potential can be written in the same form as (\ref{eq:16}) with 
\begin{subequations}
\begin{align}
 \hat{F} (\sigma) &= |\kappa|^2 \sigma^4, \\
 \hat{A} (\sigma) &= 
 \frac{1}{3} |A_\kappa| |\kappa| \sigma^3, \\
\hat{m}^2 (\sigma) &=
m_{H_2}^2 + m_S^2 \sigma^2 + m_{\tilde{L}}^2 
 + (m_{\tilde{Q}}^2 + m_{\tilde{d}_R}^2 + m_{\tilde{L}}^2) 
\frac{\lambda}{|Y_d|}\sigma.
\end{align}
\end{subequations}
By repeating the same procedure, the extremal value is obtained by
\begin{align}
|H_2|_\mathrm{ext} =
\frac{|A_\kappa|}{4 \sigma |\kappa|} \times 
\left[
1 \! + \!  \sqrt{
1 \! - \!  \frac{
8\left\{m_{H_2}^2 \!\! + \! m_{\tilde{L}}^2 \!\! + \!  m_S^2 \sigma^2 
+ \frac{\lambda \sigma}{|Y_d|}(m_{\tilde{Q}}^2 \! 
+ \!  m_{\tilde{d}_R}^2 \!\!\! + \! m_{\tilde{L}}^2)
\right\}
}{|A_\kappa|^2 \sigma^2}
}~
\right],
\end{align}
and the minimum of the potential is estimated roughly 
\begin{equation}
 V_\mathrm{min} \sim
- \frac{1}{384}\frac{|A_\kappa|^4}{|\kappa|^2} .
\end{equation}
Thus, it can be deeper than the realistic minimum if $|\kappa| \ll 1$.
A necessary condition to avoid this minimum is given by
\begin{align}
 |A_\kappa|^2
\leq 9 \left[
(m_{\tilde{Q}}^2 + m_{\tilde{d}_R}^2 + m_{\tilde{L}}^2) 
\frac{\lambda}{|Y_d|}\frac{1}{\sigma} 
+ (m_{H_2}^2 + m_{\tilde{L}}^2)\frac{1}{\sigma^2} 
+ m_S^2
\right].\label{eq:30}
\end{align}
The stringent constraint is obtained by minimizing the right-hand side of (\ref{eq:30}).
The extremal value of $\sigma$ is 
\begin{align}
 \sigma_{\mathrm ext} =
- \frac{|Y_d|}{\lambda}
\frac{2 (m_{H_2}^2 + m_{\tilde{L}}^2)}
{m_{\tilde{Q}}^2 + m_{\tilde{d}_R}^2 + m_{\tilde{L}}^2},
\end{align}
and it reads the stringent constraint
\begin{align}
\begin{split}
 |A_\kappa|^2
\leq 9 \Big[
\frac{(m_{\tilde{Q}}^2 + m_{\tilde{d}_R}^2 + m_{\tilde{L}}^2)^2}
{4 (m_{H_2}^2 + m_{\tilde{L}}^2)}
\frac{\lambda^2}{|Y_d|^2}
-2 (m_{H_2}^2 + m_{\tilde{L}}^2) + m_S^2 
\Big].
\end{split}
\end{align}

%%%%%%%%%%%%%%%%%%%%%%%%%%%%%%%%%%%%%%%%%%%%%%%%%%%%%%%%%%%%
\subsection{MSSM CCB-1 direction with gauge singlet }\label{sec:mssm-ccb-1}
%%%%%%%%%%%%%%%%%%%%%%%%%%%%%%%%%%%%%%%%%%%%%%%%%%%%%%%%%%%%
The MSSM CCB-1 with the gauge singlet direction is the direction along 
\begin{align}
 S,~H_2,~\tilde{Q},~\tilde{u}_R,~\tilde{L} \neq 0.
\end{align}
The $D$-term potential along this direction is given by
\begin{subequations}
\begin{align}
 V_D
&= \frac{1}{6} g_3^2 
(|\tilde{u}_L|^2 + |\tilde{u}_R|^2)^2 + \frac{1}{8}g_2^2
(|\tilde{u}_L|^2 + |\tilde{L}|^2 - |H_2|^2) \nonumber \\
&\quad + \frac{1}{8}g_1^2 
(\frac{1}{3}|\tilde{u}_L|^2 
-\frac{4}{3}|\tilde{u}_R|^2
-|\tilde{L}|^2
+|H_2|^2)^2.
\end{align}
\end{subequations}
The minimum becomes deeper when the $D$-term vanishes. Parameterizing the vevs as 
\begin{align}
\begin{split}
& |\tilde{u}_L| = \alpha |H_2|, \quad
 |\tilde{u}_R| = \beta |H_2|, \\
& |\tilde{L}| = \gamma_L |H_2|, \quad
 |S| = \sigma |H_2|,
\end{split}
\end{align}
the $D$-term is vanishing when 
\begin{align}
 \alpha^2 + \gamma_L^2 -1 =0, \quad
 \alpha = \beta.
\end{align}
Along this direction, the potential is given by (\ref{eq:16}) with 
\begin{subequations}
\begin{align}
 \hat{F}(\alpha, \sigma) &=
\sigma^2 (|\lambda|^2 + \sigma^2 |\kappa|^2)
+ |Y_u|^2 \alpha^2 (\alpha^2 + 2), \\
 \hat{A}(\alpha, \sigma) &=
\alpha^2 |A_u||Y_u| + \frac{1}{3}\sigma^3 |\kappa| |A_\kappa|, \\
\hat{m}^2(\alpha, \sigma) &=
 m_{H_2}^2 + \alpha^2 
(m_{\tilde{Q}}^2 + m_{\tilde{u}_R}^2)
+ \sigma^2 m_S^2 + (1 - \alpha^2) \hat{m}_{\tilde{L}}^2.
\end{align}
\end{subequations}
The constraint to avoid this minimum is obtained as 
\begin{align}
 \left(
\alpha^2 |A_u||Y_u| + \frac{1}{3}\sigma^3 |\kappa| |A_\kappa|
\right)^2 & \leq 
\left[
\sigma^2 (|\lambda|^2 + \sigma^2 |\kappa|^2)
+ |Y_u|^2 \alpha^2 (\alpha^2 + 2) 
\right] \nonumber \\
& \quad \times \left[
 m_{H_2}^2 + \alpha^2 
(m_{\tilde{Q}}^2 + m_{\tilde{u}_R}^2)
+ \sigma^2 m_S^2 + (1 - \alpha^2) \hat{m}_{\tilde{L}}^2
\right].
\end{align}
This condition can not be solved analytically and hence the extremal values of $\alpha$ and $\gamma$ 
should be determined numerically.

\bibliographystyle{apsrev}
\bibliography{biblio}

\end{document}